\documentclass[conference]{IEEEtran}
\IEEEoverridecommandlockouts

\usepackage{cite}
\usepackage{amsmath,amssymb,amsfonts}
\usepackage{algorithmic}
\usepackage{graphicx}
\usepackage{textcomp}
\usepackage{xcolor}
\def\BibTeX{{\rm B\kern-.05em{\sc i\kern-.025em b}\kern-.08em
    T\kern-.1667em\lower.7ex\hbox{E}\kern-.125emX}}
    
\usepackage[most]{tcolorbox}
\usepackage{booktabs}
\usepackage{hyperref}
\usepackage{caption}
\usepackage{subcaption}
\usepackage{xcolor}
\usepackage{listings}
\usepackage{enumitem}
\usepackage{microtype}

\usepackage[most]{tcolorbox}
\usepackage{tikz}
\usetikzlibrary{arrows.meta,positioning}
\usepackage{xcolor}

\newtcolorbox{rqbox}{enhanced, breakable=true, colback=gray!7, colframe=gray!55,
  boxrule=0.4pt, arc=1pt, left=3pt, right=3pt, top=2pt, bottom=2pt,
  before skip=6pt, after skip=4pt, fontupper=\small}

\lstdefinelanguage{pyish}{
  morekeywords={def,if,return,for,in,not,and,or,is,None},
  sensitive=true,
  morecomment=[l]{\#},
}
\lstdefinelanguage{verusrs}{
  morekeywords={fn,proof,requires,ensures,decreases,invariant,spec,open,pub,
    mod,use,struct,impl,if,else,let,return,forall,exists,assume,admit,
    reveal_with_fuel,verus,match,while,mut},
  sensitive=true,
  morecomment=[l]{//},
  morestring=[b]",
}
\definecolor{delred}{rgb}{0.99,0.87,0.87}
\definecolor{addgreen}{rgb}{0.85,0.94,0.85}

\makeatletter
\newcommand\SVparstyle[1]{{\normalfont\normalsize\bfseries\def\@IEEEsectpunct{.\ }#1}}
\renewcommand\paragraph{\@startsection{paragraph}{4}{\z@}%
  {0.8ex plus 0.3ex minus 0.2ex}{-0.5em}{\SVparstyle}}
\makeatother

\newcommand{\Proved}{\textsf{Proved}}
\newcommand{\Admitted}{\textsf{Admitted}}
\newcommand{\Hidden}{\textsf{Hidden}}

\newcommand{\severus}{\textsc{SeVerus}}
\newcommand{\severusbenchlite}{\textsc{SeVerusBench-Strix}}

\newtoggle{showrevisions}
\togglefalse{showrevisions} 

\title{\severus: Context-Aware, Gamified Benchmarking for Rust Verification}

\author{
\IEEEauthorblockN{%
  Álvaro F. Silva$^{1,\dagger}$,
  Martim Afonso$^{2,\dagger}$,
  Nuno Saavedra$^{2}$,
  Emily First$^{3}$,
  Alexandra Mendes$^{1}$,
  João F. Ferreira$^{4}$%
}

\IEEEauthorblockA{$^{1}$
Faculty of Engineering, University of Porto, Porto, Portugal\\
  \{up201603524@up.pt, alexandra@archimendes.com\}}

\IEEEauthorblockA{$^{2}$INESC-ID \& IST, University of Lisbon, Lisbon, Portugal\\
  \{martim.afonso, nuno.saavedra\}@tecnico.ulisboa.pt}

\IEEEauthorblockA{$^{3}$Rutgers University, New Jersey, USA\\
  emily.first@rutgers.edu}

\IEEEauthorblockA{$^{4}$INESC-ID and Faculty of Engineering, University of Porto, Porto, Portugal\\
  joao@joaoff.com}

\thanks{$^{\dagger}$These authors contributed equally.}
}

\date{Draft---\today}

\begin{document}

\maketitle

\begin{abstract}

Coding agents achieve high success rates on existing benchmarks for Verus, a verifier for Rust programs. These benchmarks typically provide the auxiliary lemmas needed to complete a proof. In practice, however, such lemmas must be discovered and proved. To capture this challenge, we present \severus{}, a context-aware technique to generate Verus benchmarks that organizes verification tasks into \emph{quests}, each divided into \emph{levels} in a gamified manner. Using the original proof dependencies in verified projects, \severus{} constructs one quest per verification target and progressively removes proof context to create individual levels. These levels range from the easiest, requiring completion of a single missing lemma, to the hardest, requiring a proof of the target without any supporting lemmas. An independent checker rejects changes to executable code, weakened specifications, and unsound proof constructs. A noisy binary search estimates the highest level that an agent can solve reliably.

Applied to six open-source Verus projects, \severus{} generates 819 quests comprising 10{,}095 levels. Experiments with Codex using GPT-6 Sol and GPT-6 Luna, and with Claude Code using Sonnet~4.6, show that success decreases and cost increases as proof context is withdrawn. The strongest agent solves the final level of 52 of 60 sampled quests, yet can only reliably solve 35\% of their levels, with failures concentrated in longer quests. Successful GPT-6 Sol proofs rarely reproduce the withheld lemmas, using instead alternative proof decompositions. By systematically varying proof context, \severus{} provides a foundation for constructing verification benchmarks that can continue to challenge increasingly capable coding agents.
\end{abstract}


\section{Introduction}
Coding agents generate complex software with limited human
intervention~\cite{kwa2026measuring}, but establishing that their programs satisfy
the intended requirements remains challenging~\cite{wang2026solved,yu2025utboost}.
Formal verification offers strong correctness guarantees, but constructing proofs
often requires substantial human effort~\cite{klein2009sel4,chen2025atmosphere,leroy2016compcert, rdbms, staples2014productivity, fisher2017hacms, d2008survey}.
Automating proof construction is therefore an important step towards enabling coding agents to produce verified software~\cite{10.1109/ICSE55347.2025.00161,tu2026agentic}.
This challenge is relevant in systems programming, where correctness is critical and where Rust has emerged as an increasingly popular language~\cite{jetbrains2025developer}, due in part to its emphasis on memory safety and performance. 
However, memory safety alone does not guarantee that a program satisfies its intended functional requirements~\cite{gaher2024refinedrust}.
Verus addresses this challenge by extending Rust~\cite{rust} with deductive verification, allowing developers to specify and prove functional properties of their programs~\cite{verus}. Although Verus uses SMT solvers~\cite{smt} to automatically discharge verification conditions, complex proofs often require additional guidance, including loop invariants, intermediate assertions, and auxiliary lemmas. These proof artifacts may depend on other lemmas, giving rise to intricate proof dependencies that must be resolved to verify a program.

Existing studies have evaluated the ability of coding agents to generate proofs for Verus programs. In particular, VeruSAGE-Bench~\cite{verusage} introduces a system-verification benchmark comprising 849 proof tasks extracted from eight open-source Verus repositories. The benchmark's results demonstrate the strong capabilities of modern coding agents, with the best configuration in the original study solving more than 80\% of the tasks. More recent models have reportedly achieved even higher success rates, with Claude Opus 4.5 solving 93.9\% of tasks at an average cost of \$13.65 per task~\cite{verusage_leaderboard}. 
While these results demonstrate substantial progress in automated proof generation, they leave open an important question: \emph{to what extent do existing benchmarks capture the challenges that verification agents face when auxiliary proof artifacts have not yet been developed?}

VeruSAGE-Bench does not fully capture the development environment in which proof-generation agents would operate in real-world projects. First, each task is self-contained in a single file that includes the target lemma, 
the auxiliary lemmas declarations it uses, and the specification functions and type definitions they refer to.
This setup effectively assumes that agents already know which lemmas are relevant and can retrieve all necessary context trivially, abstracting away the challenges of navigating a repository and identifying relevant proof artifacts. 
Second, and more importantly, existing tasks assume that supporting lemmas have already been established. However, in practice, verifying a program may require developing multiple auxiliary lemmas that can themselves depend on further lemmas. Consequently, proving a target property may involve constructing an entire network of supporting proofs rather than completing a single isolated proof.

We introduce \severus, a context-aware technique for
\emph{gamified benchmarking} of Rust verification agents.
We define \emph{severification} as the process of transforming an existing verified project into progressively harder verification tasks by systematically withdrawing its original proof context. An existing verified Rust project is \textit{severified} by extracting dependencies among proof artifacts to frame evaluation as a game of proof construction.
Each verification target defines a \emph{quest}, organized into successive \emph{levels} that progressively remove parts of the original proofs. Levels range from completing a single missing lemma to proving the target without its original supporting lemmas. 
Beyond completing a proof, the higher levels of a quest test whether an agent can decompose a target into intermediate lemmas, discharge the
resulting proof obligations, and retrieve the relevant definitions
and facts from a large repository.

Applied to six open-source Verus projects, \severus\ produces \severusbenchlite, an initial benchmark of 819 quests and 10{,}095 levels.
We evaluate Codex with GPT-6 Sol and GPT-6 Luna. One experiment also includes Claude Code with Sonnet~4.6.
For all three models, success decreases and cost increases as proof context is withdrawn.
Under a \$5 budget per attempt, the strongest configuration
solves the final level of 52 of 60 sampled quests, but reliably solves only 
35\% of their levels, with failures concentrated in longer quests. Successful
GPT-6 Sol proofs rarely reproduce the withheld lemmas, instead using alternative
proof decompositions.

\paragraph{Research Questions}
We address four research questions:

\def\RQOne{
How does the availability of proof context affect agents' verification success and cost?}
\def\RQOneShort{Proof Context and Solvability}
\def\RQTwo{
To what extent can \severus\ generate challenging verification tasks?}
\def\RQTwoShort{Benchmark Difficulty}
\def\RQThree{
How do agent-generated proofs differ from the original proofs?}
\def\RQThreeShort{
Agent vs. Original Proofs
}
\def\RQFour{
How does agent behavior vary with the availability of proof context?}
\def\RQFourShort{Agent Behavior and Generated Proofs}

\begin{description}
    \item[\textbf{RQ1:}] \RQOne
    \item[\textbf{RQ2:}] \RQTwo
    \item[\textbf{RQ3:}] \RQThree
    \item[\textbf{RQ4:}] \RQFour 
\end{description}

\paragraph{Contributions}
The main contributions of this paper are:
\begin{itemize}
    \item \severus, a new 
    context-aware technique for gamified benchmarking that
    constructs verification quests with systematically varied proof support.

    \item \severusbenchlite, a benchmark constructed from six real-world repositories, comprising 819 quests and 10{,}095 levels (tasks).

    \item A cost-bounded evaluation method based on noisy binary search that estimates reliable
    progress through quests without exhaustively evaluating all their levels.

    \item An empirical evaluation using two state-of-the-art coding agents to investigate how proof context affects verification success, cost, generated proofs, and agent behavior.
\end{itemize}

We make all code, experimental scripts, and data publicly available: \url{https://doi.org/10.5281/zenodo.23894085}

\section{An Illustrative Example}
\label{sec:motivating}
\def\isSorted{\texttt{is\_sorted}}
\def\lemmaThreeEq{\texttt{lemma\_3eq\_row\_count\_eq\_greater\_3}}
\def\lemmaRepeatedEq{\texttt{lemma\_repeated\_eq\_imply\_count}}
\def\lemmaCountConcat{\texttt{lemma\_count\_concat}}

\begin{figure*}[t]
\setlength{\fboxsep}{1.5pt}
\tcbset{
  severusbox/.style={
    enhanced, boxrule=0.55pt, arc=1.4mm,
    left=1.5mm,right=1.5mm,top=1.3mm,bottom=1mm,
    colback=white,coltitle=black,
    fonttitle=\bfseries\footnotesize,
    boxed title style={boxrule=0pt,arc=1.1mm},
    attach boxed title to top left={xshift=2mm,yshift=-1.2mm},
    before upper={\parindent=0pt}
  },
  rbox/.style={severusbox,
    colframe=blue!55!black,colbacktitle=blue!9},
  lonebox/.style={severusbox,
    colframe=blue!55!black,colbacktitle=blue!9},
  ltwobox/.style={severusbox,
    colframe=orange!75!black,colbacktitle=orange!13},
  lthreebox/.style={severusbox,
    colframe=green!45!black,colbacktitle=green!10},
}
\newcommand{\codefont}{\fontsize{6.3}{7.5}\selectfont}
\newcommand{\kw}[1]{\textcolor{blue!65!black}{#1}}
\newcommand{\cm}[1]{\textcolor{black!60}{#1}}
\newcommand{\ind}[1][1]{\hspace*{#1em}}
\newcommand{\note}[1]{\par\vspace{1mm}{\scriptsize #1}}
\newsavebox{\figonegrid}\newlength{\figonegridh}
\begin{lrbox}{\figonegrid}
\begin{minipage}[t]{0.57\textwidth}\vspace{0pt}
\tcbset{
  left=0.8mm,right=0.5mm,
  gridtitle/.style={
    attach boxed title to top left={xshift=0.8mm,yshift=-1.2mm}},
}
\begin{minipage}[t]{\dimexpr0.515\linewidth-0.9mm\relax}
\begin{tcolorbox}[lonebox,gridtitle,equal height group=l12,
  title={$L_1$: three consecutive equal values},
  fonttitle=\bfseries\scriptsize]
{\ttfamily\codefont
  \kw{proof fn}\par
  {\fontsize{6}{7.5}\selectfont
    lemma\_3eq\_row\_count\_eq\_greater\_3(...)}\par
  \ind\kw{requires} \cm{/* three consecutive}\par
  \ind\phantom{requires /* }\cm{equal elements */}\par
  \ind\kw{ensures} count(v, v[i]) >= 3\par}
\end{tcolorbox}
\end{minipage}\hfill
\begin{minipage}[t]{\dimexpr0.485\linewidth-0.9mm\relax}
\begin{tcolorbox}[ltwobox,gridtitle,equal height group=l12,
  title={$L_2$: no value occurs more than twice},
  fonttitle=\bfseries\scriptsize]
{\ttfamily\codefont
  \kw{proof fn}\par
  {\fontsize{6}{7.5}\selectfont
    lemma\_repeated\_eq\_imply\_count(...)}\par
  \ind\kw{requires} \cm{/* sorted, with no}\par
  \ind\phantom{requires /* }\cm{three consecutive}\par
  \ind\phantom{requires /* }\cm{equal elements */}\par
  \ind\kw{ensures} count(v, v[w]) <= 2\par}
\end{tcolorbox}
\end{minipage}\par\vspace{-1.2mm}
\begin{minipage}[t]{\dimexpr0.515\linewidth-0.9mm\relax}
\begin{tcolorbox}[lthreebox,gridtitle,equal height group=l3g,
  title={$L_3$: shared supporting lemma}]
{\ttfamily\codefont
  \kw{proof fn}\par
  lemma\_count\_concat(a, b, k)\par
  \ind\kw{ensures} count(a + b, k) ==\par
  \ind[3]count(a, k) + count(b, k)\par}
\end{tcolorbox}
\end{minipage}\hfill
\begin{minipage}[t]{\dimexpr0.485\linewidth-0.9mm\relax}
\begin{tcolorbox}[rbox,gridtitle,equal height group=l3g,
  title={Dependency graph},halign=center,valign=center]
\begin{tikzpicture}[>=Latex, font=\footnotesize,
  sym/.style={draw=black!60,rounded corners=2pt,
    minimum width=7mm,minimum height=4.4mm,
    align=center,inner sep=1.5pt},
  every path/.style={draw=black!65,line width=.65pt}]
\node[sym,fill=blue!9] (R) at (0,0) {$R$};
\node[sym,fill=blue!9] (L1) at (0.95,0.3) {$L_1$};
\node[sym,fill=orange!13] (L2) at (0.95,-0.3) {$L_2$};
\node[sym,dashed,fill=black!3,font=\scriptsize,
  minimum width=5mm] (omit) at (2.2,-0.3) {13 lemmas};
\node[sym,fill=green!10] (L3) at (3.45,0) {$L_3$};
\draw[->] (R) -- (L1);
\draw[->] (R) -- (L2);
\draw[->] (L1) -- (L3);
\draw[->,dashed] (L2) -- (omit);
\draw[->,dashed] (omit) -- (L3);
\end{tikzpicture}
\end{tcolorbox}
\end{minipage}
\end{minipage}
\end{lrbox}
\setlength{\figonegridh}{\dimexpr\ht\figonegrid+\dp\figonegrid\relax}
\centering
\begin{minipage}[t]{0.42\textwidth}\vspace{0pt}
\begin{tcolorbox}[rbox,height=0.92\figonegridh,valign=center,
  title={$R$: executable function \texttt{is\_sorted}}]
{\ttfamily\codefont 
  \kw{fn} is\_sorted(v: Vec<i32>) -> (ret: bool)\par
  \ind\kw{ensures} \cm{/* sorted \&\& each value}\par
  \ind\phantom{ensures /* }\cm{occurs at most twice */} \{\par
  \ind\kw{if} v[i-2] == v[i-1] \&\& v[i-1] == v[i] \{\par
  \ind[2]\kw{proof} \{\par
  \ind[3]\colorbox{blue!10}{\strut
    lemma\_3eq\_row\_count\_eq\_greater\_3(...);}%
    \,{\normalfont$L_1$}\par
  \ind[2]\} \kw{return} false; \}\ \cm{// ...}\par
  \ind\kw{assert} forall|i: int| ... count(...) <= 2 \kw{by} \{\par
  \ind[2]\colorbox{orange!15}{\strut
    lemma\_repeated\_eq\_imply\_count(...);}%
    \,{\normalfont$L_2$}\par
  \ind\} \kw{true} \}\par}
\end{tcolorbox}
\end{minipage}\hfill
\usebox{\figonegrid}
\caption{Simplified proof structure of \texttt{is\_sorted}.
The executable root $R$ uses $L_1$ and $L_2$, which both depend,
directly or indirectly, on the shared lemma $L_3$. Ellipses mark
omitted code, and contracts are shown schematically
(\href{https://github.com/secure-foundations/human-eval-verus/blob/main/tasks/human_eval_126.rs}{complete
file}).}
\label{fig:motivating-proof-structure}
\end{figure*}
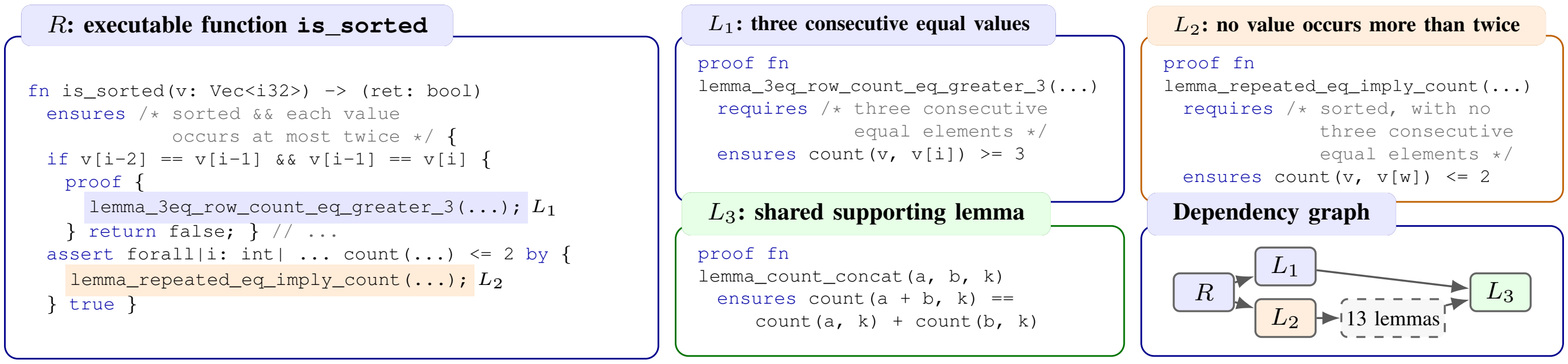

As a running example, we consider the function \isSorted,
the reference solution to HumanEval/126 from the \href{https://github.com/secure-foundations/human-eval-verus/blob/main/tasks/human_eval_126.rs}{secure-foundations}\footnote{\url{https://github.com/secure-foundations/human-eval-verus/blob/main/tasks/human_eval_126.rs}} repository (Fig.~\ref{fig:motivating-proof-structure}, $R$ box). Given a vector of integers, the function determines whether its elements are sorted in ascending order and no value occurs more than twice. Its Verus specification expresses these requirements through two universally quantified properties. 
Although the implementation is simple, its verification relies on 17 auxiliary lemmas, some depending on further lemmas. Without them, an agent must discover, state, and prove the lemmas itself. We show three of them.

\subsection{Proof dependencies}
Consider two lemmas directly used by \isSorted. The first,
\lemmaThreeEq\ ($L1$ box in Fig.~\ref{fig:motivating-proof-structure}), states that three consecutive equal elements imply that the corresponding value occurs at least three times. The second, \lemmaRepeatedEq\ ($L2$ box in Fig.~\ref{fig:motivating-proof-structure}), states that, in a sorted sequence without three consecutive equal elements, each value occurs at most twice.

Both lemmas depend, directly or indirectly, on \lemmaCountConcat\ ($L3$ box in Fig.~\ref{fig:motivating-proof-structure}), which establishes that counting the occurrences of an element distributes over sequence concatenation. 
Thus, removing $L3$ affects multiple proof obligations, illustrating how the difficulty of verifying a function depends on the number of missing lemmas and also on their relationships.
The second lemma additionally relies on a larger subgraph of supporting lemmas.
The dependency graph in Fig.~\ref{fig:motivating-proof-structure} illustrates these relationships.
An edge from one symbol to another indicates that the former invokes the latter.

\subsection{A quest of verification tasks}
\severus\ introduces the idea of \emph{gamified benchmarking} for Rust verification. For each verification target, it constructs a \emph{quest}: a sequence of verification challenges called \emph{levels}, each providing progressively less of the original proof context. 
This allows us to evaluate not only whether an agent can verify a target, but also how much proof support it requires.
At each level, the verification target and its supporting lemmas can be in one of three states, illustrated in Fig.~2: \emph{proved} (green), meaning the original proof is retained; \emph{admitted} (orange), meaning the statement is retained but its proof is removed to be reconstructed; or \emph{hidden} (dashed), meaning the lemma is removed entirely, requiring the agent to discover its own auxiliary reasoning.

\begin{figure*}
    \centering
    \includegraphics[width=\linewidth]{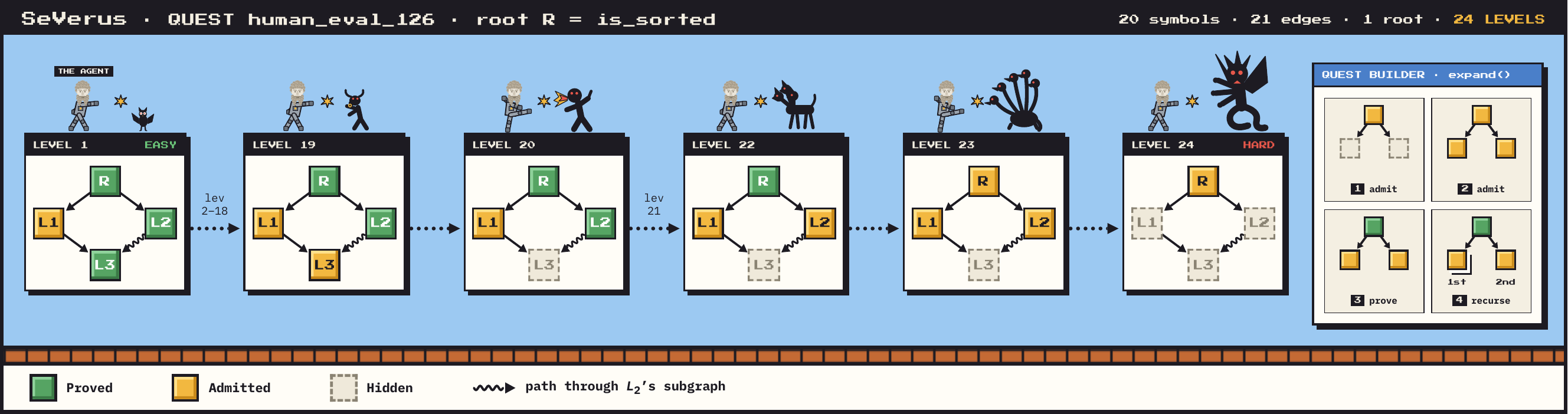}
    \caption{The quest of \texttt{is\_sorted}: six of its 24 levels (left) and the four steps of \texttt{expand()} that generate it (right).}
    \label{fig:quest}
\end{figure*}

Fig.~\ref{fig:quest} illustrates six of the 24 levels generated for the \texttt{is\_sorted} function from Fig.~\ref{fig:motivating-proof-structure}. At Level 1, nearly all the original proof context is available, leaving only a single lemma for the agent to prove. As levels increase, more proof support is withdrawn. By Level 22, both lemmas directly supporting the executable function are admitted, while their shared supporting lemma is hidden. At Level 23, the executable function itself also becomes a proof obligation. Finally, at Level 24, the agent must verify the executable function without its original proof annotations or supporting lemmas. This final level is called the \emph{root level}. 
In general, a proof is removed exactly when its corresponding lemma or function is admitted or hidden.
In this verification game, a root is the \emph{final boss}; one needs to be an expert in proof decomposition and obligation dispatching to conquer it. The levels before it are intermediate bosses: each removes a different part of the original proof, so the agent must prove, rediscover, or rely on a different set of lemmas.

\section{The \severus\ Benchmarking Framework}
\label{sec:formalism}
We describe how \severus\ generates the levels of a quest, how a submission is graded, and how an agent is scored.

\subsection{Task Generation} 

\paragraph{Dependency Graph Construction}
\label{sec:corpus}

Given a verified Verus project, \severus\ constructs a directed dependency graph $G=(V,E)$, where $V$ 
contains the symbols of the project's executable, proof, and specification functions, and $E \subseteq V \times V$ represents dependencies between them. An edge $(u,v)\in E$ indicates that $u$ invokes $v$. Each symbol has a unique identifier of the form \texttt{<file>:<line>:<name>} and a source span.

We further distinguish a subset $P \subseteq V$ of proof functions and lemmas, which represent explicit proof obligations. 
For the example of Fig.~\ref{fig:motivating-proof-structure}, $|V|=20$ and $|E|=21$.

\paragraph{Identification of Verification Targets}
\label{sec:roots-verification-targets}
\severus\ identifies verification targets by analyzing the
roots of the dependency graph. 
A root is a symbol on which no other symbol depends (i.e., no other symbol invokes it).
Formally,
given a dependency graph $G=(V,E)$, a symbol $r \in V$
is a \emph{root} if it has no incoming edges, i.e.,
$\nexists u \in V : (u,r) \in E$.
Each 
root is the verification target of one quest. 
Throughout the paper, we refer to a quest's root as its target. For projects with multiple roots, \severus\ generates an independent quest for each.

We call the set of symbols reachable from a symbol in $G$ its \emph{cone}. 
To construct a quest, \severus\ considers only the cone of its root.
In the running example (Fig.~\ref{fig:motivating-proof-structure}),
the executable function \texttt{is\_sorted}, denoted by $R$,
is the sole root, as no other symbol depends on it.

The cone of $R$ contains $L_1$, $L_2$, and their supporting lemmas.

\paragraph{Proof Context Transformations}
\label{sec:proof-context-transformations}

To generate verification levels, \severus\ systematically
varies the availability of proof artifacts while preserving
the executable behavior and correctness requirements of the
original program.

\severus{} assigns each symbol one of three states: \Proved, \Admitted, or \Hidden.
Each state determines how the corresponding symbol is
represented in the generated level:

\begin{itemize}[leftmargin=*]
    \item \textbf{\Proved.}
    The symbol retains its original definition, including
    its implementation and proof annotations. The agent
    can rely on it without reproving it.

    \item \textbf{\Admitted.}
    The symbol retains its original signature and
    specification, but the agent must supply its removed proof.

    \item \textbf{\Hidden.}
    The symbol is removed entirely from the generated
    project, with its specification and proof.
\end{itemize}
The admitted state is implemented differently for proof
functions and executable functions.
For a proof function, \severus\ replaces its body with an \texttt{assume(false)} stub, so other symbols can still reference the lemma. The stub is an obligation the agent must discharge.
For an executable function, \severus\ removes its
proof annotations, including proof blocks, assertions,
loop invariants, and decreases clauses, while preserving
its executable statements and specification.
Executable functions are never hidden, since doing so
would remove the program being verified.

\paragraph{Quest Generation}
\label{sec:quest-generation}
For each root identified, \severus\
constructs a quest (\S\ref{sec:roots-verification-targets}). The process starts with the root task, in which the
target's proof annotations are removed and its supporting
lemmas are hidden. \severus\ then progressively restores
the original proof context through two state transitions:
\emph{admit}, which changes a hidden symbol to admitted,
and \emph{prove}, which changes an admitted symbol to
proved. Each transformation step produces a new task,
or \emph{level}, of the quest. A single step may admit
multiple symbols simultaneously.

\paragraph{Dependency-aware expansion}
Quest generation follows a depth-first traversal of the
root's cone. For each symbol, \severus\ first
admits the symbol and its immediate dependencies,
then restores the symbol's original proof, and finally
recursively expands its dependencies. 
Dependencies are
processed in decreasing order of cone size, so
dependencies with larger cones are explored first. 
This deterministic order restores the largest
withheld cone first, so a dependency shared by several cones is restored inside the largest of them. 
The right panel of Fig.~\ref{fig:quest} shows the expansion process on a symbol with two dependencies.

\paragraph{Level ordering}
Levels are numbered in the reverse order of generation. For a quest with $n$ levels, Level~1 gives the most original proof support and Level~$n$ is the root task, with the least. This order reflects the available proof context, not necessarily difficulty, which RQ1 investigates empirically.

\paragraph{Obligations of a level} At every level of a quest, the target is the root. The \emph{obligations} of a level are all Admitted symbols in the root's cone, including the root itself. A level is solved when all of them are proved. At Level~1, the root is \Proved{} and a single lemma is \Admitted{}, so that lemma is the only obligation. The final level is the one where only the root is \Admitted{} and all other dependencies are \Hidden{}.

\paragraph{Illustrative example}
For the running example in Fig.~\ref{fig:motivating-proof-structure}, \severus\ generates
a quest comprising 24 levels.
The traversal expands $L2$ before $L1$, since
$L2$ has a larger cone.
Consequently, the shared lemma $L3$ is restored
through $L2$'s dependencies before the traversal
reaches $L1$. Since $L3$ is already proved,
it is not processed again. Fig.~\ref{fig:quest} shows the states of the four named symbols at selected levels.

\subsection{Agent Evaluation}
\label{sec:scoring}
\label{sec:rq2-method}
\label{sec:task-evaluation}

An attempt on a \emph{level} 
is considered \emph{solved} only
if the agent's submission passes both \emph{verification} and \emph{our anti-cheat}: \texttt{make verify} runs Verus on the project with the agent's changes, and \texttt{make anticheat} guarantees soundness by ensuring that the agent has not modified the executable code, weakened the specifications, or introduced unsound shortcuts. Agents may add auxiliary lemmas and proof annotations, provided they verify.
In general, attempting every level of every quest is expensive: quests reach hundreds of levels, and an
attempt may run out of budget. 
Also, because agents are nondeterministic, a single attempt is insufficient to reliably assess success near the boundary between solvable and unsolvable levels.
To account for this variability, \severus\ evaluates agents using a noisy binary search~\cite{karp2007noisy}, which assumes that difficulty increases across levels, an assumption empirically supported by RQ1.

\paragraph{Noisy binary search}
Let $\psi(\ell)$ denote the probability that an agent solves level~$\ell$ in a single attempt. 
Our goal is to estimate $\ell^* = \max\{\ell : \psi(\ell) > 0.5\}$, the highest level at which the agent succeeds more often than it fails.
If $\psi$ is non-increasing across levels, $\ell^*$ also gives the number of levels with success probability above $0.5$. To estimate this threshold, \severus\ uses a noisy binary search. At each step, it runs up to three attempts
on a level $\ell$
and determines whether $\psi(\ell) > 0.5$ by majority vote. Each step halves the remaining search interval, requiring at most $3\lceil\log_2(n+1)\rceil$ attempts for a quest of $n$ levels. The search narrows the estimated threshold to a single level.

\paragraph{Benchmark score} For a quest $q$ of $n_q$ levels, $\ell^*_q = n_q$ when the 
agent solves the root level.
Simply averaging the per-quest scores $\ell^*_q/n_q$ gives equal weight to quests of different lengths. Since short quests are both more numerous and often solved completely, they would disproportionately influence the overall score. Instead, \severus\ weights each quest by its number of levels, computing the Reliable Level Score
$RLS =  \sum_q \ell^*_q \,/\, \sum_q n_q$.
Under the assumed difficulty ordering, $RLS$ estimates the proportion of benchmark levels that the agent solves reliably. This weighting treats each level as a distinct verification task: longer quests contribute more, much like conventional benchmarks in which projects contribute different numbers of tasks.

\section{Experimental Methodology}
\label{sec:evaluation}

\subsection{Benchmark Construction and Characterization}

\paragraph{Projects}
We \textit{severify} six open-source Verus projects that cover different kinds of systems software to build \severusbenchlite\ (Table~\ref{tab:dataset}).

We selected these projects because they are among the largest open-source Verus-verified systems, including Anvil, the largest~\cite{verusage}. All but TLA-RS appear in VeruSAGE-Bench~\cite{verusage}. Of its other three projects, Memory Allocator and Node Replication do not verify with any Verus release we could build, and Atmosphere needs a custom Nix-based multi-crate build, which we leave for future work.

\paragraph{Generated quests} \severus\ generates one quest per root
(\S\ref{sec:quest-generation}). 
\textit{Severifying} the six projects yields 819 quests with 10,095 levels in total (Table~\ref{tab:dataset}).
Most quests are short. In five projects the median quest
has three to five levels, because most roots depend on only a few lemmas. A few quests are
long, and they account for most of the levels.

\begin{table}[t]
  \centering
\caption{The \severusbenchlite\ benchmark. 
\emph{Full benchmark}: Rust code size (kLOC), proof functions, quests (one per target), levels (one task each), and median/maximum quest length. \emph{Sample}: total levels across the ten RQ2 quests per project, and minimum/maximum agent attempts required by noisy binary search.}
  \label{tab:dataset}
  \footnotesize
  \setlength{\tabcolsep}{2.5pt}
  \resizebox{\columnwidth}{!}{
\begin{tabular}{l rrrrr rr}
\toprule
 & \multicolumn{5}{c}{\textbf{Full benchmark}} & \multicolumn{2}{c}{\textbf{Sample}} \\
\cmidrule(lr){2-6} \cmidrule(l){7-8}
\textbf{Project} & \textbf{kLOC} & \textbf{Proofs} & \textbf{Quests} & \textbf{Levels} & \textbf{Length} & \textbf{Levels} & \textbf{Attempts} \\
\midrule
Vest~\cite{vest} & 10.9 & 221 & 147 & 544 & 3 / 11 & 68 & 50--96 \\
IronKV~\cite{ironfleet, verus_case_studies} & 10.9 & 91 & 96 & 356 & 3 / 13 & 82 & 54--105 \\
PMemLog~\cite{verus_case_studies} & 3.4 & 45 & 16 & 167 & 8.5 / 35 & 157 & 72--129 \\
NRKernel~\cite{verus_case_studies} & 26.6 & 289 & 82 & 3{,}953 & 5 / 214 & 968 & 100--180 \\
TLA-RS~\cite{tla_rs_method} & 94.6 & 1{,}447 & 427 & 3{,}935 & 3 / 237 & 528 & 86--159 \\
Anvil~\cite{anvil} & 83.9 & 600 & 51 & 1{,}140 & 3 / 894 & 981 & 64--123 \\
\midrule
\textbf{All} & 230.3 & 2{,}693 & 819 & 10{,}095 & & 2{,}784 & 426--792 \\
\bottomrule
\end{tabular}
}
\end{table}

\paragraph{Evaluated sample} Running every level of every quest is not affordable, so
we evaluate a sample of ten quests per project.
For each project, we first take its longest quest, which RQ1 also uses. We then draw nine
further quests at random with a fixed seed among the quests that have at least five levels,
until the project has ten. 
The sample contains 2{,}784 levels, representing 28\% of all benchmark levels. 
Long quests dominate the sample, meaning that most evaluated levels correspond to the projects' main theorems.
The last column of Table~\ref{tab:dataset} shows that the search methodology used in RQ2 (\S\ref{sec:rq2-method}) requires 426--792 attempts on the sample, compared with 10{,}095 attempts to evaluate every benchmark level once.

\paragraph{The longest quests} 
The root of each longest quest corresponds to a verification target with a large number of dependencies, typically representing a key correctness property of the project. At the root level, the agent must rebuild every proof in the root's cone.
On NRKernel (214 levels), the root \texttt{unmap\_aux} removes a virtual address from the x86 page table and proves that the abstract map loses exactly that address while the page-table invariants still hold. On Anvil (894 levels), the root \texttt{core\_holds} proves that four Kubernetes controllers sharing one cluster are safe and eventually reach a stable reconciliation, a liveness property.
Our artifact~\cite{severusartifact} describes the longest quests of the other four projects.

\subsection{Agents and Execution Environment}
\label{sec:execution-environment}

\paragraph{Agents} We evaluate Codex~\cite{openai_codex} with GPT-6 Sol and GPT-6 Luna, and, in RQ1, also Claude
Code~\cite{anthropic_claude_code} with Claude Sonnet~4.6.
We initially evaluated Sonnet 4.6, the latest version available through our API of the model used by VeruSAGE, the state-of-the-art system for automated Verus proof generation~\cite{verusage}, alongside GPT-6 Sol, a state-of-the-art model at the time of writing. Since GPT-6 Sol substantially outperformed Sonnet 4.6, we selected it for the subsequent experiments. We also evaluated GPT-6 Luna as a lower-cost alternative to assess the performance of a more cost-effective model.
Each attempt starts in a fresh container with a fresh agent
context and stops at a cost of \$5 or after four hours. The task description tells the agent to keep working until the environment ends the attempt. A level is solved if the final submission passes both
verification and the anti-cheat (\S\ref{sec:task-evaluation}).

\paragraph{Task packaging} Each task is a Docker image whose workspace holds the modified
sources, a \texttt{Makefile} with the \texttt{verify} and \texttt{anticheat} targets, and a task
description.
Original sources and grading metadata never enter the agent's container.
\texttt{make anticheat} is an independent Rust checker that compares the submission with the
original project. It requires the non-ghost code of executable functions to be byte-identical,
every \texttt{requires} and \texttt{ensures} clause to be unchanged, and the target's dependency
scope, computed from the call graph of the submission so that helpers the agent adds are
included, to contain no \texttt{assume(...)}, \texttt{admit()},
\texttt{\#[verifier::external\_body]} or \texttt{\#[verifier::admit]}.

\paragraph{Access to the original proofs} As the source projects are public, an agent with
internet access might download the proof it is asked to rebuild. A preliminary GPT-6 Sol
campaign did exactly this: in 11 attempts on NRKernel and Anvil, it cloned the original
repository and copied its proofs. We discarded that campaign and ran Sol and Luna with the
network restricted: the container can reach only the model API, web search is disabled, and the
task description states that there is no internet access. Most Sonnet attempts predate the
restriction; we checked their logs, and none used the network or read the grading metadata.

\section{Experimental Results}

RQ1 tests whether quest levels are ordered by difficulty, the assumption behind \severus's scoring (\S\ref{sec:scoring}).
RQ2 applies this scoring method to 60 quests to assess agent performance.
RQ3 compares the proofs of the solved attempts of RQ1 and RQ2 with the original proofs,
and RQ4 examines how the agents produced them.

\subsection{RQ1: \RQOne}
\label{sec:rq1}

\paragraph{Method}
\label{sec:rq1-method}
All levels of a quest share the target and its specification. They differ only in how much of the
original proof they retain.
Although higher levels are expected to be harder, difficulty does not necessarily increase monotonically. An \Admitted{} lemma used by a \Proved{} proof must be proved as stated, whereas a \Hidden{} lemma may be replaced by an alternative. For example, in Fig.~\ref{fig:quest}, if $L_1$ is difficult but a simpler lemma would also suffice to prove $R$, Level~22 forces the agent to prove $L_1$. At the root level (Level~24), however, the agent may use the simpler lemma instead, potentially making the final boss easier than Level~22.

We test the null hypothesis that the probability $\psi(\ell)$ of solving Level~$\ell$ remains constant across a quest against the alternative that it decreases as the level increases. We evaluate Claude Code with Sonnet~4.6 and Codex with GPT-6 Sol and GPT-6 Luna on the longest quest of each project, ranging from 11 levels on Vest to 894 on Anvil. For each quest, we select 35 equally spaced levels in advance (or all levels if fewer) and make one attempt at each, starting at Level~1 and progressing toward the root. After four consecutive failures, we switch to six equally spaced levels between the current level and the root, including the root.
Once an agent stops succeeding, higher levels are also expected to fail, and failed attempts tend to be the most expensive. This expectation held for 40 of the 42 attempts made after a switch, allowing us to save 160 attempts across the three long quests.
For NRKernel, TLA-RS, and Anvil, the grid samples only part of each quest, so we can locate where success stops only within the grid spacing of 6, 7, and 26 levels, respectively. A trend on the sample reflects across the quest: the grid is fixed before any attempts, and only the switch depends on the outcomes, which the statistical test described below accounts for. The experiment comprised 332 attempts at a total cost of \$597: 93 with Sonnet (\$333), 115 with Sol (\$218), and 124 with Luna (\$46).

For each quest, we test for a decrease in success across levels using the one-sided Mann--Kendall test~\cite{mann1945nonparametric,kendall1948rank}. Since the switch makes subsequent attempted levels outcome-dependent, we use $\alpha=0.025$ and validate through simulations that the false-positive rate remains below 5\% (see artifact ~\cite{severusartifact}). We aggregate the six quests per agent using the stratified Mann--Kendall test and compare agents using an exact McNemar test~\cite{mcnemar1947note} on commonly attempted levels. To assess cost trends, we compute one-sided Kendall's $\tau_b$~\cite{kendall1945treatment} between level and cost, both across all attempts and solved attempts only, as Sonnet and Sol failures almost always incur the full \$5 cap.

\begin{figure*}[t]
  \centering
  \includegraphics[width=\textwidth]{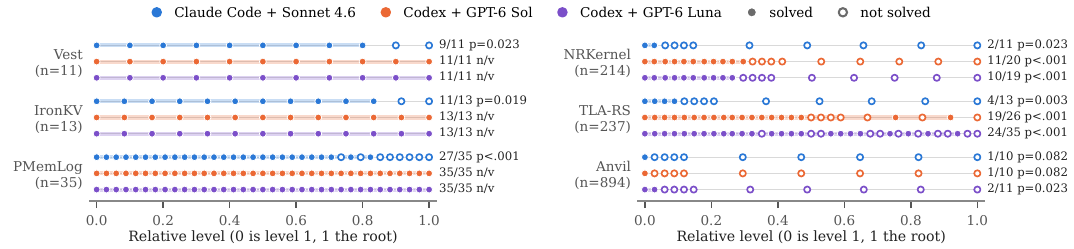}
  \caption{Outcome of each attempt by relative level (0 = Level~1, 1 = root). Filled markers indicate success. Right-hand labels report successful/total attempts and Mann--Kendall $p$-values (n/v: no variance, as all attempts succeeded).}
  \label{fig:rq1-trace}
\end{figure*}

\begin{table}[t]
\caption{
RQ1 per quest of $n$ levels. Bold: furthest agent; ${}^{*}$: significant success decrease; $\tau_b$ on solved attempts (--: under three). \emph{Pooled}: stratified $Z$ for success decrease and cost increase.
}
\centering
\footnotesize
\setlength{\tabcolsep}{2pt}
\begin{tabular}{lrrrrrrrrr}
\toprule
 & \multicolumn{3}{c}{\textbf{Highest solved}} & \multicolumn{3}{c}{\textbf{$\tau_b$(level, cost)}} & \multicolumn{3}{c}{\textbf{Cost (\$)}} \\
\cmidrule(lr){2-4}\cmidrule(lr){5-7}\cmidrule(lr){8-10}
\textbf{Project ($n$)} & \textbf{Sonnet} & \textbf{Sol} & \textbf{Luna} & \textbf{Sonnet} & \textbf{Sol} & \textbf{Luna} & \textbf{Sonnet} & \textbf{Sol} & \textbf{Luna} \\
\midrule
Vest (11) & 9$^*$ & \textbf{11}\phantom{$^*$} & \textbf{11}\phantom{$^*$} & $.72$\phantom{$^*$} & $.75$\phantom{$^*$} & $.49$\phantom{$^*$} & 36 & 6.3 & 0.3 \\
IronKV (13) & 11$^*$ & \textbf{13}\phantom{$^*$} & \textbf{13}\phantom{$^*$} & $.89$\phantom{$^*$} & $.79$\phantom{$^*$} & $.67$\phantom{$^*$} & 29 & 8.2 & 0.8 \\
PMemLog (35) & 29$^*$ & \textbf{35}\phantom{$^*$} & \textbf{35}\phantom{$^*$} & $.69$\phantom{$^*$} & $.62$\phantom{$^*$} & $.70$\phantom{$^*$} & 115 & 20 & 2.7 \\
NRKernel (214) & 7$^*$ & \textbf{64}$^*$ & 57$^*$ & --\phantom{$^*$} & $.71$\phantom{$^*$} & $.82$\phantom{$^*$} & 50 & 74 & 16 \\
TLA-RS (237) & 22$^*$ & 218$^*$ & \textbf{223}$^*$ & $1.00$\phantom{$^*$} & $.82$\phantom{$^*$} & $.83$\phantom{$^*$} & 54 & 67 & 18 \\
Anvil (894) & 1\phantom{$^*$} & 1\phantom{$^*$} & \textbf{27}$^*$ & --\phantom{$^*$} & --\phantom{$^*$} & --\phantom{$^*$} & 48 & 43 & 8.0 \\
\midrule
\textbf{Pooled} ($Z$) & $5.7$$^*$ & $5.0$$^*$ & $4.6$$^*$ & $6.4$$^*$ & $8.2$$^*$ & $9.0$$^*$ &  &  &  \\
\textbf{Total} &  &  &  &  &  &  & 333 & 218 & 46 \\
\bottomrule
\end{tabular}

\label{tab:rq1}
\end{table}

\paragraph{Success falls and cost rises with the level}
\label{sec:rq1-results}
The pooled row of Table~\ref{tab:rq1} combines the six projects of each agent into one
stratified test, which uses every attempt and is therefore our strongest evidence. For all three
agents success falls and cost rises as the level rises, and cost rises also when only solved
attempts are counted (all $p<.001$). 

Fig.~\ref{fig:rq1-trace} shows a similar pattern across almost all quests: agents solve the shallower levels but begin to fail beyond a certain depth. Successes and failures rarely intermingle, resulting in a generally sharp transition. The Mann--Kendall test confirms a decreasing success trend for most quests with varying outcomes (Table~\ref{tab:rq1}). 
The only exception is Anvil, where no statistically significant trend is detected due to the sparse grid, which results in too few attempts per agent near the transition point to reliably detect a performance decline.
TLA-RS is the only quest where this transition is not sharp.
Sol solves every level up to 112, fails the next
four, and then still solves two of the five levels it attempts after
the switch, up to level 218; Luna alternates between successes
and failures from level 84 up to level 223,
so the highest solved level overstates where success stops, which motivates the
repeated attempts of the search used in RQ2. We discuss why TLA-RS behaves
differently in \S\ref{sec:rq2-tla}.

The agents reach different failure points, with Sonnet never surpassing the GPT-6 agents. On the three small projects, Sonnet fails before the root, while Sol and Luna solve every level. On NRKernel, Sol reaches Level~64 and Luna Level~57; both reach high levels on TLA-RS, but only Luna progresses beyond Level~1 on Anvil. Paired comparisons confirm this ordering: Sol and Luna each solve about 20 levels that Sonnet fails, while Sonnet solves none that either fails (exact McNemar,
$p<10^{-5}$). Sol and Luna are statistically indistinguishable.

At its highest solved level, an agent proves the target and all supporting
lemmas not provided at that level. On the three small projects, Sol and
Luna rebuild the entire proof, and Sonnet most of it. On the larger
projects, all agents rebuild only part: Sol and Luna prove about 160 of
180 lemmas on TLA-RS, 47 and 42 of 153 on NRKernel, respectively, while
on Anvil (655 lemmas), they prove the target only when almost the entire
proof is provided. Since these quests are sampled, these counts are
lower bounds.

Cost increases with level for all three agents on every project.
For Sonnet and Sol, this is partly because nearly all failures reach
the \$5 cap, but the trend also holds for successful attempts alone
on quests with sufficient data ($\tau_b$ in Table~\ref{tab:rq1}).
Luna costs about an order of magnitude less than Sol at every level,
while Sol is several times cheaper than Sonnet where both succeed.
Sonnet and Sol exhaust nearly the full budget on successes just below
the switch, so their failure points depend partly on the \$5 cap.
Luna never reaches the cap, so its failures cannot be attributed to
the budget.

\begin{rqbox}
\textbf{Answer to RQ1:} 
As proof context is withdrawn, success decreases and cost increases for all three agents (stratified tests, $p<.001$). Quest levels thus form a generally increasing, though not strictly monotonic, difficulty scale, with agents reaching different depths.
\end{rqbox}

\subsection{RQ2: \RQTwo}
\label{sec:rq2}

\paragraph{Method} We run the noisy binary search (\S\ref{sec:scoring}) with GPT-6 Sol and
GPT-6 Luna on the 60 quests of the evaluated sample, under the same
conditions as in RQ1, and score each agent with $RLS$.
The search skips most levels, so we first check it on the longest quest of each project. We
also estimate $\ell^*$ from the RQ1 attempts, with the isotonic-regression method of Oron and
Flournoy~\cite{Oron03072017}, and compare the two.

\paragraph{The search agrees with RQ1 grid}
\label{sec:rq2-results}

On the longest quests, search-derived $\ell^*$ agrees with the RQ1 grid estimate for all 12 agent--quest pairs: both find the root solved on the three small projects, and all estimates for the long quests fall within the grid's 95\% intervals. Search requires fewer attempts at similar cost (76 vs.\ 115 for Sol; 80 vs.\ 124 for Luna) while pinpointing $\ell^*$ to a single level rather than intervals spanning tens of levels. It places Sol at levels 62/214 (NRKernel), 104/237 (TLA-RS), and 23/894 (Anvil), and Luna at 64, 108, and 27, respectively. The full
comparison is in our artifact~\cite{severusartifact}.

\begin{table}[t]
\caption{RQ2 results over the ten quests of each project. \emph{Root}: quests solved up to the
root. \emph{Mean}: per-quest mean of $\ell^*/n$. \emph{Attempts}: agent attempts the search made.}
\label{tab:rq2}
\centering
\footnotesize
\setlength{\tabcolsep}{3pt}
\begin{tabular}{lrrrrrrrrrr}
\toprule
 & \multicolumn{2}{c}{\textbf{Root}} & \multicolumn{2}{c}{\textbf{Mean}} & \multicolumn{2}{c}{\textbf{$RLS$}} & \multicolumn{2}{c}{\textbf{Attempts}} & \multicolumn{2}{c}{\textbf{Cost (\$)}} \\
\cmidrule(lr){2-3}\cmidrule(lr){4-5}\cmidrule(lr){6-7}\cmidrule(lr){8-9}\cmidrule(lr){10-11}
\textbf{Project} & \textbf{Sol} & \textbf{Luna} & \textbf{Sol} & \textbf{Luna} & \textbf{Sol} & \textbf{Luna} & \textbf{Sol} & \textbf{Luna} & \textbf{Sol} & \textbf{Luna} \\
\midrule
Vest     & 10 & 10 & 1.00 & 1.00 & 1.00 & 1.00 & 50  & 50  & 16  & 1.7 \\
IronKV   & 10 & 10 & 1.00 & 1.00 & 1.00 & 1.00 & 55  & 55  & 32  & 6.4 \\
PMemLog  & 10 & 10 & 1.00 & 1.00 & 1.00 & 1.00 & 72  & 72  & 26  & 4.9 \\
NRKernel &  5 &  4 & .58  & .55  & .22  & .22  & 113 & 116 & 372 & 91  \\
TLA-RS   &  8 &  6 & .89  & .87  & .67  & .65  & 92  & 102 & 181 & 38  \\
Anvil    &  9 &  9 & .90  & .90  & .11  & .12  & 69  & 68  & 140 & 20  \\
\midrule
\textbf{All}      & 52 & 49 & .90  & .89  & .35  & .35  & 451 & 463 & 768 & 162 \\
\bottomrule
\end{tabular}
\end{table}

At the quest level, Sol appears close to solving the benchmark,
reaching the root in 52 of 60 quests. At the level granularity,
however, it reliably solves only 35\% of the levels ($RLS=0.35$).
Sol solves every quest with fewer than 50 levels up to the root,
and Luna all but two. Failures are concentrated in longer quests:
all eight quests that Sol leaves unsolved have at least 86 levels.
Luna performs nearly as well as Sol, solving 49 quests up to the root versus Sol's 52. The three additional failures occur on NRKernel and TLA-RS. Sol reaches a higher level on 6 quests,
Luna on 5, and the remaining are ties. Both achieve the same $RLS=0.35$.

\paragraph{Why TLA-RS behaves differently}
\label{sec:rq2-tla}

TLA-RS quests are as long as NRKernel's, yet agents solve most of their levels (RLS 0.67 and 0.65, against 0.22), and their results are the least stable. Success still decreases significantly with the level (Table II), but slowly: over a wide range of levels $\psi(\ell)$ appears to stay close to one half, so attempts flip between success and failure, and RQ1 still succeeds at levels 218 (Sol) and 223 (Luna). Estimating $\ell^*$ from the RQ1 attempts with a 95\% confidence interval (details in the supplementary material of our artifact~\cite{severusartifact}) gives an interval of about 100 levels on TLA-RS. The search's $\ell^*$ (104 and 108) falls inside it, but with at most three attempts per probe the search cannot separate success probabilities that are too close across levels. The reason is that from level 112 to 223 every level asks for the same three inductive lemmas of the Raft safety proof. These lemmas are likely to dominate the success probability as every failure of Luna in this range fails on exactly these three, while the added lemmas are proved.

\begin{rqbox}
\textbf{Answer to RQ2:} The strongest agent, GPT-6 Sol, solves 52 of 60 quests up to the root
but only about a third of their levels ($RLS=0.35$). GPT-6 Luna scores the same 
at a fifth of the cost. Almost all unsolved levels lie on the long
quests of NRKernel, TLA-RS and Anvil. 
Results show that \severus\ generates challenging tasks even for the strongest agents.
\end{rqbox}

\subsection{RQ3: \RQThree}

\paragraph{Method}

We measure how agent-generated proofs
differ from the original solutions. While we cannot rule out
that agents helped write them, each project’s top contributors
suggest that most are at least human-revised. Only in TLA-RS, the most recent project, is an agent (Claude Code) the
main contributor, while the other projects have few agent-made
commits. We therefore report each analysis with and without
TLA-RS.

We compare the proofs across the 267 passed tasks for GPT-6 Sol, pooling the RQ1 and
RQ2 walks and counting each task once (averaging its repeated attempts).
At each
level, we compare the agent’s solution with the omitted set, i.e.,
the hidden lemmas, the admitted lemmas and the target proof.
We consider only successful attempts, since an incomplete proof
does not show whether the agent was converging towards a
correct solution.
For each task with hidden lemmas, we compare the generated lemmas against the ground truth, grouping results into four categories: no new lemmas, different lemmas, some reused, and most reused. Matching is based on lemma statements, with names removed, parameters $\alpha$-renamed, and external anchors (vstd lemmas, spec functions, and types) preserved verbatim; proof bodies are excluded. Two lemmas share a statement key if and only if their canonicalized statements match, regardless of name or proof. Solutions are compared by the intersection of their statement sets over their union:
\begin{equation*}
    \mathit{stmt\_jaccard} = \frac{\text{agent stmt} \cap \text{human stmt}}{\text{agent stmt} \cup \text{human stmt}}
\end{equation*}

\emph{Re-stated recall} measures the fraction of deleted reference statements matched by agent-generated lemmas, ignoring additional helpers. It captures reference-claim coverage and thus provides an upper bound on agreement, whereas the symmetric Jaccard also penalises extra agent lemmas.

We also compare proof dependency structures through call graphs, identifying each lemma by its statement (as defined above). Ideally, two solutions should match if one can be transformed into the other by consistently renaming their own lemmas while preserving external anchors. We approximate this equivalence using the Weisfeiler--Leman (WL) colour-refinement algorithm~\cite{weisfeiler}. Each node is initially coloured by its canonical statement, with names removed, parameters renamed, and external anchors preserved; proof bodies are discarded. We then iteratively refine node colours using the colours of the lemmas they call. Consequently, two nodes share a final colour if they express the same claim and depend on lemmas with matching colours, capturing equivalent dependency structures up to renaming. Each solution is represented by the multiset of its nodes' final colours, and structural similarity is computed as the multiset Jaccard similarity between the resulting fingerprints:
\begin{equation*}
    \mathit{stmt\_graph} = \frac{\#(\text{shared colours})}{\#(\text{colours in either})}
\end{equation*}

\noindent where both counts are taken with multiplicity.
We evaluate, within traces, how the increasing difficulty (represented by the depth) affects this similarity.

Beyond this, we examine two aspects associated with proof quality: assertion density and code duplication. For each proof, we measure how much proof code is duplicated rather than abstracted into reusable lemmas. We use a sliding window of 20 consecutive tokens, hashing each window and marking those whose normalized sequence occurs elsewhere in the solution, either within the same lemma or across different lemmas.

\begin{figure}[!htbp]
    \centering
    \includegraphics[width=\linewidth]{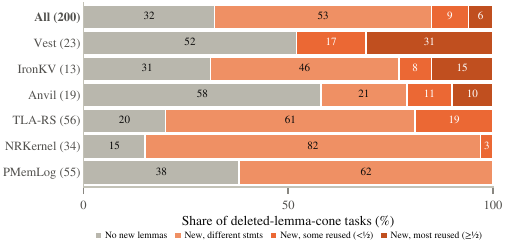}
    \caption{Lemma reuse across projects.}
    \label{fig:statement_reuse}
\end{figure}

\paragraph{Statement and Graph Comparison}
Figure~\ref{fig:statement_reuse} groups deleted-cone tasks into four outcomes: no new lemmas, different lemmas, some re-stated lemmas, or most re-stated lemmas.
Across the 200 tasks whose target depends on
deleted \Hidden\ lemmas (the \emph{All projects} bar), agents build no new lemmas in 32\%, entirely different lemmas in 53\%, and re-state at least one in only 15\%. Results vary by project: 15--58\% of tasks involve no new lemmas, while statement reuse ranges from substantial (Vest 48\%, IronKV 23\%) to virtually none (PMemLog / NRKernel $\approx 0$).

From Table~\ref{tab:stmt-recall} we see that recall is the most lenient reading, yet it is barely above the symmetric overlap (0.11 vs.\ 0.08 pooled), so the low reproduction of lemmas is genuine rather than an artifact of penalizing extra helpers; the one exception is Vest, whose agents re-state three-quarters of the deleted claims.

\begin{table}
  \centering
  \caption{Reference deleted-lemma coverage on each project's \Hidden\ tasks. \emph{Re-stated recall} measures one-sided coverage (ignoring extra agent helpers), while \Hidden-only statement \emph{Jaccard} measures symmetric overlap. GPT-6 Sol, RQ1+RQ2 pooled, one row per task.}
  \label{tab:stmt-recall}
\setlength{\tabcolsep}{3pt}
\begin{tabular}{@{}lrrrrrr|r@{}}
\toprule
 & \textbf{Vest} & \textbf{Anvil} & \textbf{IronKV} & \textbf{TLA-RS} & \textbf{NRKernel} & \textbf{PMemLog} & \textbf{All} \\
\midrule
Tasks             & 23   & 19 & 13   & 56   & 34 & 55   & 200 \\
Recall  & 0.73 & 0.11 & 0.10 & 0.02 & 0.01 & 0.00 & 0.11 \\
Jaccard & 0.56 & 0.06 & 0.05 & 0.01 & 0.00 & 0.00 & 0.08 \\
\bottomrule
\end{tabular}

\end{table}

\begin{figure}[!htbp]
  \centering
  \begin{minipage}[t]{0.485\textwidth}
    \centering
    \includegraphics[width=\linewidth]{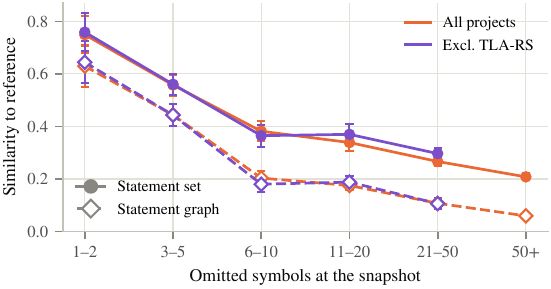}
    \caption{Variation of proof similarity with the quest depth.}
    \label{fig:rq3_similarity_depth}
  \end{minipage}\hfill
\end{figure}

Figure~\ref{fig:rq3_similarity_depth} shows $stmt\_jaccard$ and $stmt\_graph$ across quest-depth bins, including results for all projects and excluding TLA-RS, whose solutions may have been influenced by the use of a coding agent.
The similarity between agent-generated and human-written solutions declines sharply with depth: as more lemmas are withheld, the agent's solution increasingly diverges from the reference in terms of both claims and structure. The gap between the two metrics further indicates that, even when agents reuse some of the same lemmas as the reference, they often place them in a different dependency context.
Deeper levels are therefore less exposed to data
contamination.
Including TLA-RS has little effect on the overall results. 
Although it may have been influenced by a coding agent, we hypothesize that differences in the harness and model family account for the distinct characteristics of its proofs.

\paragraph{Coding Practices}
Code duplication and assertion use give insight into proof conciseness and maintainability.
Repeated proof fragments may indicate a lack of possible abstraction or reuse, while frequent assertions may reflect greater reliance on explicit intermediate facts to guide the prover. As neither metric is inherently undesirable (duplication can improve local clarity, and assertions can make proof steps more explicit), we treat them as complementary indicators of proof structure. 

\begin{figure}
    \centering
    \begin{minipage}{0.48\linewidth}
        \centering
        \includegraphics[width=\linewidth]{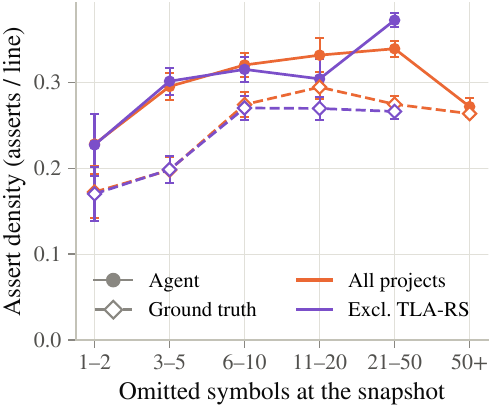}
        \caption{Assert density across depth (fraction of proof lines that are assertions).}
        \label{fig:assert_density}
    \end{minipage}
    \hfill
    \begin{minipage}{0.48\linewidth}
        \centering
        \includegraphics[width=\linewidth]{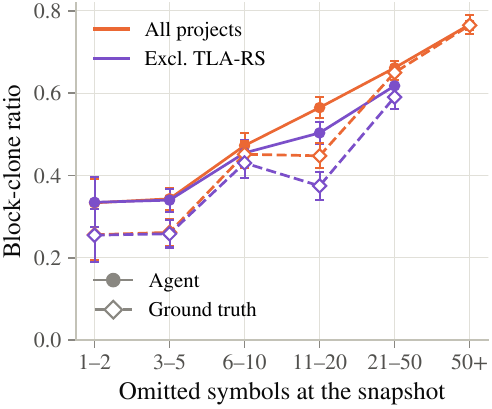}
        \caption{Clone ratio across depth (fraction of tokens in recurring 20-token windows).}
        \label{fig:code_duplication}
    \end{minipage}
\end{figure}

Figure~\ref{fig:assert_density} reports \emph{assert density}, the fraction of proof lines consisting of assertions. It increases on shallower tasks before plateauing, with agents exceeding the reference at every depth except the deepest (0.30 vs.\ 0.24 overall; higher in 71\% of tasks). As the measure is normalized by proof length, this difference cannot be attributed simply to agents writing longer proofs. It is present on the easiest levels and generally widens with depth. Including TLA-RS narrows the gap only in the deepest bin, which consists entirely of this project's levels. TLA-RS has a corpus-wide density roughly twice that of any other project, suggesting that the apparent convergence in bins reflects the project's unusually high reference assert density, rather than a change in agent behaviour.
Excluding TLA-RS, agents discharge proof obligations using a higher density of intermediate assertions than the human-written references, and this difference tends to increase with task depth.

Figure~\ref{fig:code_duplication} shows that agent-generated proofs generally have a higher block-clone ratio than human-written proofs across depth bins. This difference is absent for TLA-RS, whose commits were reportedly produced by coding agents. This suggests that agent-generated proofs tend to contain more duplicated code, although the TLA-RS results indicate that this tendency may depend on the agent and its development setup.

\begin{rqbox}
\textbf{Answer to RQ3:} 
Agent proofs differ substantially from the references: across 200 levels, they reproduce only 11\% of deleted lemma statements (re-stated recall), with just 8\% symmetric statement overlap. Structural similarity also decreases sharply with quest depth, indicating growing divergence in both lemmas used and dependency structure. Agent proofs contain more assertions and duplicated code, and this trend generally increases with depth. 
\end{rqbox}

\subsection{RQ4: \RQFour}
\label{sec:rq4}

\paragraph{Method} We take every RQ1 and RQ2
attempt of GPT-6 Sol and GPT-6 Luna with at least one lemma to rebuild (565 and 581 attempts), average the attempts of each level so that it counts once (314 and 320
levels), and count \emph{decisions}, the tool calls other than polls of a running command. A trend is the slope of
a measure on $\log_2$ of the lemmas to rebuild, with a fixed effect per project, given as a factor
per doubling. 
We record each attempt's cost, tokens, turns, thinking time, and failure outcome. For agent-terminated failures, we analyse the closing reasoning to classify why it stopped

\paragraph{Deeper levels take more steps, not larger ones} The median attempt takes 20 decisions
at 1--3 lemmas to rebuild and 157 at 64 or more for Sol, and 33 and 337 for Luna, a factor of
1.39 and 1.44 per doubling. Verification runs and edits grow similarly, with agents maintaining a consistent edit–verify loop of 0.5–0.9 edits per verification run across all depths. Most of the growth comes beyond 16 lemmas, where solve rates drop from 95\% to
70\% and 21\% for Sol, and from 94\% to 63\% and 32\% for Luna. The final diff is largely additive: net additions grow from 24 to 326 lines (Sol) and 31 to 724 (Luna), with at most 20 deletions. Agents mainly fill in proof bodies and add lemmas, rarely restructuring existing code.

\begin{table}[t]
\caption{How failed attempts end (left) and, for those the agent ends itself, why
it stops (right), from a reading of every one of these stops. S: GPT-6 Sol, L: GPT-6 Luna.}
\label{tab:rq4-endings}
\centering
\footnotesize
\setlength{\tabcolsep}{3pt}
\resizebox{\columnwidth}{!}{%
\begin{tabular}{@{}lrr@{\hspace{1.6em}}lrr@{}}
\toprule
\textbf{How it ends} & S & L & \textbf{Why the agent stops} & S & L \\
\midrule
\textbf{Failed attempts} & 107 & 126 & mid-plan, no stated reason & 5 & 44 \\
\quad \$5 budget & 88 & 0 & task impossible as stated & 3 & 17 \\
\quad 4-hour limit & 0 & 25 & work too hard, out of ideas & 4 & 25 \\
\quad agent reports it incomplete & 19 & 99 & believes time or tokens run out & 7 & 6 \\
\quad agent claims success & 0 & 2 & broke workspace, believes done & 0 & 9 \\
\bottomrule
\end{tabular}

}
\end{table}

\paragraph{Agents often stop working midway} 

Neither agent fully follows the instruction to keep working until the environment ends the attempt. As Table~\ref{tab:rq4-endings} shows, Sol largely complies: 88 of its 107 failed attempts end when the \$5 budget is exhausted, after a median of 18 minutes of reasoning. Thus, most of Sol's failures indicate that the budget was insufficient, making it part of what determines $\ell^*$. Luna voluntarily ends 101 of its 126 failed attempts, at a median cost of \$0.4; the remaining 25 run until the four-hour limit. The largest category of Luna's voluntary stops (44/101) occurs mid-plan, immediately after re-running both checks, with no stated reason. Otherwise, both agents tend to judge the task impossible as stated, either because a lemma requires a stronger precondition (17 cases for Luna, 3 for Sol) or because the remaining work seems too difficult (25 and 4 cases, respectively). These judgments are often premature: the original project proves every withdrawn lemma, and another attempt by the same agent solves the same level in 6 of Luna's 17 cases deemed impossible and all 3 of Sol's.

\paragraph{Budget in dollars buys Luna far more work} 
Sol is held back by the \$5 budget and Luna by its own stopping. Luna still gets almost as far as Sol (RQ2) because its tokens cost about 18 times less (\$0.014 against \$0.26 per million input tokens, mostly cached). On levels with 16 or more lemmas to rebuild, a solved Luna attempt takes a median of 224 turns and 48 minutes of thinking for \$0.49, against 78 turns and 11 minutes for \$1.94 for Sol. Over all attempts, Luna processes 3.8 times as many input tokens as Sol at a fifth of the cost (\$207 against \$988). On the 44 levels where a Sol attempt ran out of budget, Luna works a median of 79 minutes for \$0.73 and solves 16 of them, 6 of which no Sol attempt solves.
Under a budget in dollars, the ranking of agents therefore depends on the price per
token as much as on the model.

\paragraph{The anti-cheat steers the agents back to proving}
Tasks start with the anti-cheat failing on Admitted stubs. Thus, we only count cheats introduced by agents, across the 1,362 (Sol) and 2,298 (Luna) self-invoked anti-cheat runs and final grading. These shortcuts, accepted by the verifier alone, occur in 13\% of Sol's attempts and 24\% of Luna's. Most involve stubs:
the agent reinstates needed hidden lemmas using \texttt{assume(false)} or \texttt{external\_body} rather than proving them. Another shortcut is placing helper lemmas in a new \texttt{verus!} block, outside the code the checker permits agents to modify. Cheats involving frozen code are almost exclusively Luna's: 17 attempts edit a spec function, 5 alter the contract of a trusted function, and 5 modify an executable function. Most spec edits only add triggers or casts, but some weaken the proof obligations.
In Anvil, Luna removed the \texttt{external\_body} of the trusted lemma \texttt{vrs\_vsts\_str\_neq}, proved the first of its two facts and deleted the second; once flagged, it restored and proved both, solving the level.
Flagging pivots agents to
the actual task: 67 of Sol's 72 and 101 of Luna's 138 flagged attempts were solved, with all
shortcuts replaced by a proof, and only 3 submissions that pass verification still fail the anti-cheat.

\begin{rqbox}
\textbf{Answer to RQ4:} Deeper levels lead agents to take more small steps, not different ones. Sol mostly exhausts the \$5 budget, while Luna often stops early for ungrounded reasons. Luna's costs let it work much longer on the same budget, nearly matching Sol. The anti-cheat checker keeps the agents on the tasks: it flags a shortcut in 13\% (Sol) and 24\% (Luna) of attempts, mostly stubbed lemmas, and once flagged the agents replace most of them with proofs.
\end{rqbox}

\section{Threats to Validity}
\paragraph{Internal validity} The \$5 budget limits where success stops (RQ1). Since agents are nondeterministic, reruns may change individual results with one attempt per level (RQ1) or three per probe (RQ2), but are unlikely to alter pooled trends. Although agents may recall public proofs despite network restrictions, RQ3 shows they rarely reproduce the original lemmas. The initial task description also announced an unenforced 30-minute thinking limit, which some agents treated as binding (RQ4). We therefore reran every walk and search where Luna voluntarily ended a failed attempt, and those where such a stop could have affected Sol's result, without this limit.

\paragraph{Construct validity} Each quest uses one depth-first withdrawal order through the root's dependency cone. Other orders, such as smaller subtrees first, could produce different difficulty profiles. Evaluating multiple orders would require repeating the evaluation for each, which is prohibitively expensive.

\paragraph{External validity} All tasks come from six Verus projects verified with one Verus
release, and we evaluate two agents (three in RQ1). Other projects, verifiers or
agents may behave differently. The models are versioned products that change over time; we report
the versions we used, and include every attempt's trajectory in our artifact.

\section{Related Work}

\paragraph{Agents for Software Verification}

AutoVerus~\cite{autoverus} and VeruSAGE~\cite{verusage} employ multi-agent pipelines for proof generation and repair, incorporating verifier feedback, planning, and context management. AlphaVerus~\cite{alphaverus} translates programs from established verification languages (e.g., Dafny~\cite{dafny}) into Verus, refining them through verifier feedback and reusing verified programs as exemplars. RagVerus~\cite{ragverus} uses retrieval-augmented generation to provide repository context through semantic and dependency-based retrieval, while SAFE~\cite{chen2025automated} adopts a self-evolving approach based on synthetic proof generation, verifier feedback, and model fine-tuning.
Related work in theorem proving explores the generation and reuse of auxiliary lemmas.
LEGO-Prover~\cite{wanglego} builds a library of verified lemmas for reuse or synthesis during proof construction, while DeepSeek-Prover-V2~\cite{ren2025deepseek} and Hilbert~\cite{varambally2026hilbert} use hierarchical decomposition to break complex theorems into intermediate subgoals.
While these approaches improve proof generation, they do not systematically evaluate agent performance as proof dependencies become harder to resolve. \severus\ addresses this gap by using dependency graphs from real-world projects to construct progressively harder quests, requiring agents to generate and prove auxiliary lemmas before tackling the target theorem. This enables evaluation of both individual proof synthesis and the handling of interdependent proof obligations.

\paragraph{Benchmarks}

Existing Verus benchmarks vary in scale and scope, from isolated proof synthesis to end-to-end verification. Verus-Bench~\cite{autoverus} provides 150 algorithm-level tasks, while HumanEval-Verus~\cite{humanevalverus} adapts HumanEval into 85 function-level tasks; both require generating proof annotations for existing implementations and specifications. VeriContest~\cite{vericontest} comprises 946 competitive-programming problems and evaluates specification, code, and proof generation separately or end-to-end. Other work does this for Lean~\cite{ye2026verina, thakur2025clever}. 
VeriStruct~\cite{veristruct} targets 129 functions across 11 Rust data-structure modules, requiring generation of abstraction functions, invariants, specifications, and proofs. VeruSAGE-Bench~\cite{verusage} offers 849 proof-generation tasks from eight open-source Verus projects, covering real-world systems software. 
In all of them, each task is a standalone
unit with its proof context.

Other work explores controlled context availability: VeriSoftBench~\cite{xin2026verisoftbench} for Lean and Cobblestone~\cite{kasibatla2026cobblestone} for Rocq investigate curated proof contexts. However, neither systematically evaluates agents as auxiliary lemmas become explicit proof-generation objectives. CoqDev~\cite{lu2026adaptive}, which served as an initial inspiration for our work, increases task difficulty by masking theorems and lemmas introduced in the same commit as the target proof. However, its masking strategy follows the project's commit history rather than proof dependencies, whereas \severus\ follows the dependency graph of the proofs. 


\section{Conclusion and Future Work}
\label{sec:future}
We presented \severus, a technique that turns verified Rust projects into quests of verification
tasks that share one target and withdraw its proof context step by step, from a single missing
lemma up to the whole proof. 
\textit{Severifying} six open-source Verus projects resulted in 819 quests with 10{,}095 levels,
and it scores an agent with a noisy binary search that needs a roughlt 450 attempts to evaluate 60 traces.
attempts per project instead of one per level. The levels of a quest generally form a difficulty scale for the agents we tested. The strongest agent reaches the root of most sampled quests but solves only 35\% of their levels, and agents that succeed without the original lemmas rarely rebuild them.

\severusbenchlite{} is only the first version of the dataset. Once agents saturate it,
we will release harder versions, each named after a stronger monster, as in
Fig.~\ref{fig:quest}. Besides adding more projects, we can make the tasks harder by
removing the specification functions that the target's statement does not use, and
the \texttt{requires} and \texttt{ensures} clauses of every function except the target.
The agent would then also have to write the definitions and intermediate contracts
its proof needs. Finally, since \severus{} relies only on the dependency graph of the
proofs, it could be adapted to other verifiers. 

\begingroup
\renewcommand{\thefootnote}{}
\footnotetext{We used generative AI tools from Anthropic and OpenAI to assist with coding, text rewriting, and the generation of Figures~\ref{fig:motivating-proof-structure} and ~\ref{fig:quest}.}
\endgroup


\bibliographystyle{IEEEtran}
\bibliography{references}

\end{document}